\documentclass[twocolumn,amsmath,trackchanges]{aastex702}

\begin{document}

\title{Helium-poor winds do not require helium-poor planets }

\correspondingauthor{Leonardos Gkouvelis}

\author[orcid=0000-0002-1397-8169]{Leonardos Gkouvelis}
\affiliation{Instituto de Astrofísica de Andalucía (IAA-CSIC), Glorieta de la Astronomía s/n, E-18008 Granada, Spain}
\email[show]{gkouvelis@iaa.es}

\author[orcid=0000-0003-1572-7707]{Francisco J. Pozuelos}
\affiliation{Instituto de Astrofísica de Andalucía (IAA-CSIC), Glorieta de la Astronomía s/n, E-18008 Granada, Spain}
\email{francisco@email.address}
 
\begin{abstract}

Observations of the metastable He,{\sc i},10830,\AA\ triplet have revealed escaping exoplanet atmospheres whose inferred helium abundances range from nearly nebular compositions to strongly helium-depleted winds. Such depletion is commonly interpreted as evidence for atmospheric evolution, preferential escape, or departures from primordial composition. We present a closed-form analytic transport-retention theory for the atmospheric transition region between the homopause and the base of the planetary wind. The theory quantifies the competition between upward transport and molecular separation, demonstrating that transport physics alone can substantially deplete the helium abundance supplied to the escaping flow.  Our solution yields a retention factor, $\chi_{\rm He}$, that measures the fraction of helium supplied to the base of the hydrodynamic wind relative to the deep atmospheric abundance. Combining the analytic framework with numerical forward models of the He,{\sc i},10830,\AA\ absorption and a sample of twelve helium-observed sub-Neptunes and mini-Neptunes, we find that observed systems span the full range of predicted retention states, from nearly complete retention to strong helium depletion. These forward models use the helium abundance supplied by the analytic framework as the lower-boundary condition to predict the corresponding He,{\sc i},10830,\AA\ absorption. We further show that increasing helium retention systematically strengthens the expected absorption signal by increasing the helium reservoir available to the upper atmosphere. These results suggest that helium depletion does not necessarily imply intrinsically helium-poor atmospheres or evolutionary transitions toward secondary compositions, but may instead be a natural consequence of transport physics in escaping atmospheres.
\end{abstract}


\section{Introduction} 
\label{sec:intro}

Hydrodynamic escape is one of the most efficient atmospheric mass-loss
regimes and is expected to play a major role in the evolution of
close-in exoplanets (\citealt{Watson1981,Yelle2004,Murray2009,Owen2019}).  The most successful tracer of this process to date is the metastable
He\,{\sc i}\,10830\,\AA\ triplet
(\citealt{Nortmann2018,Spake2018,Kirk2020,Orell2024}),
which can be routinely observed from the ground and benefits from a
well-developed theoretical framework for forward modelling
(\citealt{Oklopvcic2018,Linssen2022,DosSantos2022,Lampon2023}).
Recent surveys have revealed a diverse population of escaping
atmospheres, including strong detections, weak signals, and
non-detections across hot Jupiters, warm Neptunes, sub-Neptunes, and
mini-Neptunes.

A recurring result of hydrodynamic and multi-species escape models is
that the helium abundance in the escaping flow is often significantly
lower than expected from a primordial H/He atmosphere. For example,
\citet{Xing2023} showed that mass fractionation can reproduce the low
He/H ratio inferred for HD~209458b without requiring a globally
helium-poor envelope, while \citet{Taylor2026} found that diffusive
separation may strongly suppress helium in the high-gravity planet
HD~149026b. At the same time, weak helium signals and helium depletion
are frequently interpreted as evidence for atmospheric evolution or
transitions away from primordial compositions
(\citealt{Kobayashi2026}). This raises a fundamental question: to what
extent does the composition of the escaping wind actually reflect the
composition of the deep atmosphere?

At the population level, helium observability also follows broad
empirical trends involving the incident XUV flux and planetary gravity.
\citet{Zhang2022} related the observed equivalent width to the amount of
occulting metastable helium material, while \citet{Forcada2025} found a
positive correlation between helium absorption strength and an
energy-limited escape proxy. More recently, \citet{Allan2025} showed
that both observed helium equivalent widths and detailed hydrodynamic
model predictions broadly follow these empirical trends, although with
significant scatter and uncertainties associated with stellar XUV
luminosities, stellar winds, and assumed lower-boundary He/H ratios.
These studies suggest that atmospheric escape is a key ingredient in
helium observability, but they do not directly address how the helium
abundance supplied to the escaping flow is established.

Previous multi-fluid hydrodynamic studies have shown that diffusive fractionation can naturally produce helium-depleted escaping atmospheres without requiring helium-poor planetary envelopes (e.g., \citet{Xing2023,Schulik2025,Taylor2026}). However, these studies solve the coupled transport equations numerically, leaving the governing transport dependencies to be inferred from the simulations. In this work, we investigate the transport processes operating between the homopause and the base of the hydrodynamic wind. We derive a closed-form analytic transport-retention framework for this transition region, providing a theoretical description to numerical models. Our solution predicts the fraction of helium reaching the base of the escaping flow relative to the deep atmospheric abundance. We then explore the consequences of this framework for primordial
H/He atmospheres and for the interpretation of helium observations.
Using numerical escape calculations, we illustrate how variations in
helium retention affect the expected He\,{\sc i}\,10830\,\AA\ signal.
Finally, we apply the model to the helium-observed sample of
\citet{Allan2025}, combining observational constraints and
multi-species hydrodynamic simulations to estimate the retention state
of each planet. We show that substantial helium depletion can arise
naturally from transport physics alone, even while the deep atmosphere
retains a primordial composition.

\section{Vertical transport} \label{sec:transport}

\begin{figure}
\centering
\includegraphics[width=\columnwidth]{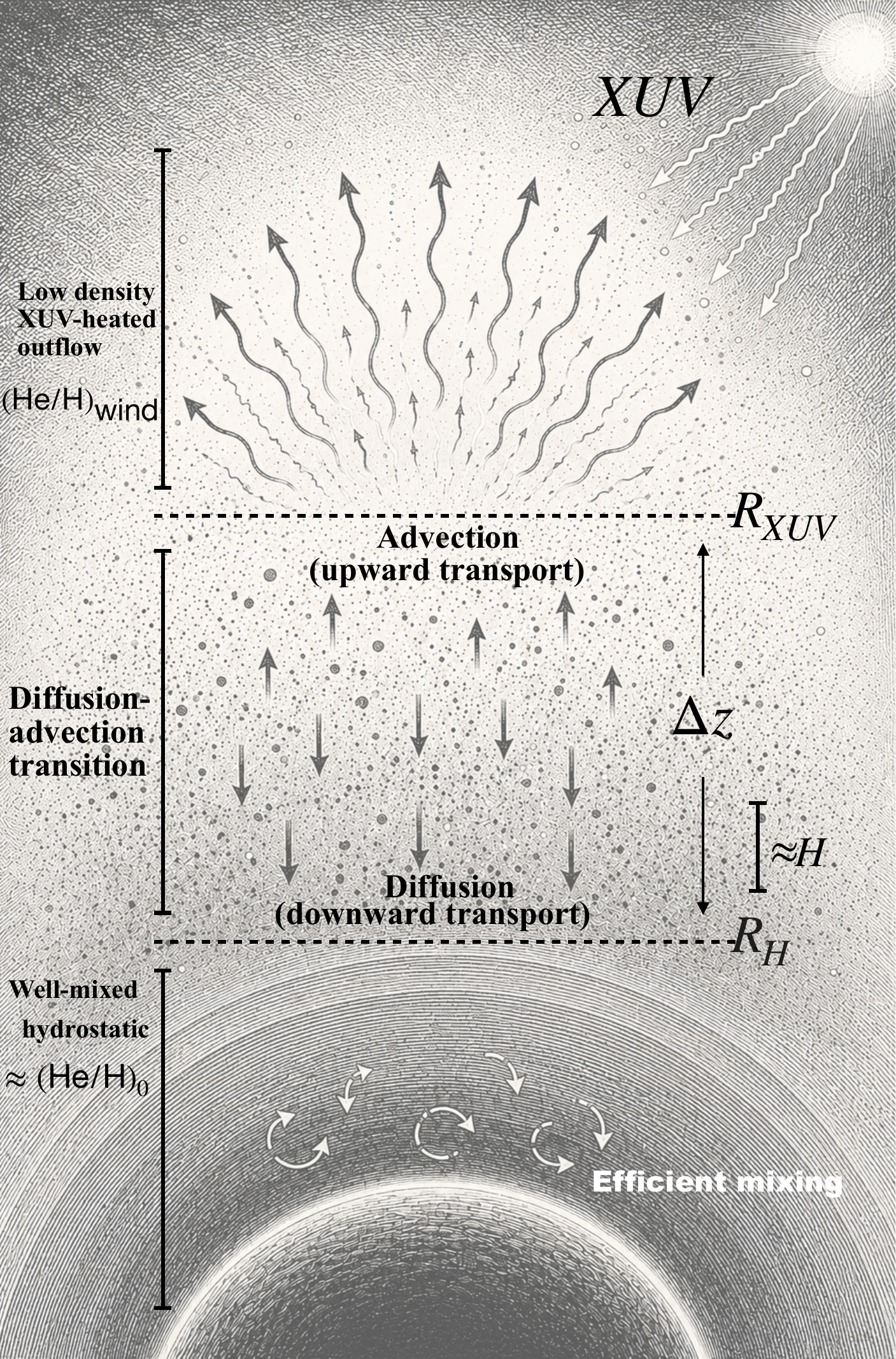}
\caption{
Schematic representation of the atmospheric transition region between the homopause and the base of the hydrodynamic wind. Below the homopause, turbulent mixing maintains a well-mixed atmosphere. Above the XUV absorption layer, the atmosphere is in hydrodynamic outflow. Between these boundaries lies the transport region, where upward advection competes with molecular diffusion and species separation. The helium retention factor derived in this work quantifies how these competing processes modify the composition supplied to the escaping flow.}
\label{fig:cartoon}
\end{figure}

The composition of the escaping wind is determined by transport
processes operating between the well-mixed lower atmosphere and the
base of the planetary wind where the hydrodynamic outflow is initiated. We
consider the transition region bounded below by the homopause, where
eddy mixing and molecular diffusion are comparable
($K_{zz}\simeq D$), and above by the XUV absorption layer, defined
approximately by $\tau_{\rm XUV}\simeq1$. The atmospheric composition
at this upper boundary provides the lower boundary condition for the
escaping wind. Understanding how species are transported across this
transition region is therefore essential for predicting the composition
of the outflow.

For a minor species \(i\) transported through a dominant background species \(j\), the steady-state vertical flux can be written as  
\begin{equation}
\Phi_i
=
\underbrace{n_i u(z)}_{\rm Advection}
-
\underbrace{
D_{ij} n_j
\frac{d f_i}{dz}
}_{\rm Concentration\ diffusion}
-
\underbrace{
D_{ij} n_i
\left(
\frac{1}{H_i}
-
\frac{1}{H_j}
\right)
}_{\rm Gravitational\ separation},
\end{equation}

where \(f_i=n_i/n_j\), \(u(z)\) is the bulk upward velocity, \(D_{ij}\) is the binary diffusion coefficient, and \(H_i\) and \(H_j\) are the scale heights of species \(i\) and \(j\), respectively (\citet{Banks1973}).

Substituting \(n_i=f_i n_j\) and dividing by \(D_{ij}n_i\) yields

\begin{equation}
\frac{d\ln f_i}{dz}
=
\frac{u(z)}{D_{ij}}
-
\left(
\frac{1}{H_i}
-
\frac{1}{H_j}
\right)
-
\frac{\Phi_i}
     {D_{ij}n_i}.
\label{eq:general_transport}
\end{equation}

Equation (\ref{eq:general_transport}) represents the exact steady-state balance between hydrodynamic transport, molecular separation, and the finite upward flux of the species itself.

The gravitational-separation term can be rewritten as

\begin{equation}
\frac{1}{H_i}
-
\frac{1}{H_j}
=
\frac{(m_i-m_j)g}
{k_{\rm B}T}
=
\alpha_{ij}
\frac{1}{H_j},
\end{equation}

where

\begin{equation}
\alpha_{ij}
=
\frac{m_i-m_j}{m_j}.
\label{eq:open_mass}
\end{equation}

We define the dimensionless thickness of the transition region, shown in Figure \ref{fig:cartoon}, as

\begin{equation}
\Lambda
=
\frac{\Delta z}{H_j}.
\end{equation}

Within the quasi-hydrostatic transition layer, the background density approximately follows

\begin{equation}
n_j(z)
=
n_{j,0}
\exp
\left(
-\frac{z}{H_j}
\right).
\end{equation}

Assuming that the upward particle flux of the dominant constituent is approximately conserved,

\begin{equation}
n_j(z)u(z)
\simeq
{\rm const},
\end{equation}

the velocity profile becomes

\begin{equation}
u(z)
=
u_b
\exp
\left[
-\frac{\Delta z-z}{H_j}
\right],
\end{equation}

where \(u_b\) denotes the velocity at the wind base. 

Integrating Equation (\ref{eq:general_transport}) from the lower transition region to the wind base yields

\begin{equation}
\ln \chi_i
=
-\alpha_{ij}\Lambda
+
\int_0^{\Delta z}
\frac{u(z)}{D_{ij}}
\,dz
-
\int_0^{\Delta z}
\frac{\Phi_i}
     {D_{ij}n_i}
\,dz,
\label{eq:chi_exact}
\end{equation}

where $\chi_i
=
f_{i,b}/
     f_{i,0} $ is the retention factor of species \(i\). The final term in Equation~(\ref{eq:chi_exact}) represents the finite
escape flux of the minor constituent. Writing
$\Phi_i=n_iw_i$, where $w_i$ is the drift velocity of species $i$,
gives

\begin{equation}
\frac{\Phi_i}{D_{ij}n_i}
=
\frac{w_i}{D_{ij}}.
\end{equation}

In the homopause--wind base  region, the atmospheric mass flux is
expected to be dominated by the escaping hydrogen background, such that
the drift velocity of the minor species remains small compared with the
bulk flow velocity ($w_i\ll u$). The finite-species-flux term therefore
represents a higher-order correction and is neglected in the following
analysis.

With this leading-order approximation, the abundance profile is controlled by the competition between hydrodynamic transport and gravitational separation.

The advective contribution then becomes

\begin{equation}
\int_0^{\Delta z}
\frac{u(z)}{D_{ij}}
\,dz
=
\frac{u_bH_j}{D_{ij}}
\left(
1-e^{-\Lambda}
\right).
\end{equation}

Defining

\begin{equation}
\Pi_{ij}
=
\frac{u_bH_j}{D_{ij}},
\end{equation}

the retention factor becomes

\begin{equation}
\boxed{
\chi_i
=
\exp
\left[
-\alpha_{ij}\Lambda
+
\Pi_{ij}
\left(
1-e^{-\Lambda}
\right)
\right].
}
\label{eq:chi_general}
\end{equation}
Equation~(\ref{eq:chi_general}) provides a general solution for the
retention of a minor constituent transported through a dominant
background gas. The solution depends on two dimensionless parameters:
the atmospheric extent $\Lambda$, which measures the geometric
thickness of the transition region in units of scale height, and the
transport efficiency $\Pi_{ij}$, which quantifies the competition
between upward advection and molecular diffusion. 

The transport parameter $\Pi_{ij}$ is closely related to the classical
crossover-mass formalism of \citet{Hunten1987}. In a hydrodynamically
escaping atmosphere, the crossover mass is

\begin{equation}
m_c
=
m_j
+
\frac{k_{\rm B}T\,F_j}
     {b_{ij}g\,n_j},
\end{equation}

where $F_j$ is the upward particle flux of the dominant species,
$b_{ij}$ is the binary diffusion parameter, and $m_j$ is the mass of
the background gas. Using $F_j=n_ju$, $D_{ij}=b_{ij}/n_j$, and
$H_j=k_{\rm B}T/(m_jg)$ yields

\begin{equation}
m_c
=
m_j
\left(
1+
\frac{uH_j}{D_{ij}}
\right),
\end{equation}

such that

\begin{equation}
\Pi_{ij}
=
\frac{uH_j}{D_{ij}}
=
\frac{m_c-m_j}{m_j}.
\end{equation}

This relation demonstrates that $\Pi_{ij}$ is effectively a
dimensionless crossover-mass excess. Large values of $\Pi_{ij}$
correspond to efficient hydrodynamic entrainment and strong
compositional coupling, whereas small values correspond to
diffusion-dominated transport. Unlike the classical crossover-mass
criterion, which is fundamentally local, Equation
(\ref{eq:chi_general}) describes the integrated competition between
transport and molecular separation across the entire homopause--XUV
transition region through its combined dependence on $\Pi_{ij}$ and
$\Lambda$.

\section{Applications to primordial atmospheres}
\label{sec:applications}

We first apply the analytic framework to atmospheres of primordial
composition. This class includes giant planets, sub-Neptunes, and many
mini-Neptunes for which helium escape observations are now available.
Moreover, primordial H/He atmospheres provide the simplest application
of the model because helium is transported through a hydrogen-dominated
background, avoiding the additional complexity of fully multi-species
transport (\citealt{Gkouvelis2025}).

It follows from \ref{eq:open_mass}, that for nebular composition $\alpha_{{\rm He},{\rm H}}\simeq
3$ so that the helium retention factor, equation \ref{eq:chi_general},  becomes

\begin{equation}
\chi_{\rm He}
=
\exp
\left[
-3\Lambda
+
\Pi_{{\rm He},{\rm H}}
\left(
1-e^{-\Lambda}
\right)
\right].
\label{eq:chi_he_final}
\end{equation}
Equation~(\ref{eq:chi_he_final}) predicts that helium retention is
controlled by the competition between two effects: gravitational
separation, represented by the term $3\Lambda$, and upward
hydrodynamic transport, represented by $\Pi_{\rm He,H}$. The balance
between these processes determines whether the escaping flow remains
compositionally coupled to the deep atmosphere or becomes depleted in
helium before reaching the wind base.

\begin{figure*}
\centering
\includegraphics[scale=0.26, trim=0cm 2.5cm 0cm 2cm, clip]{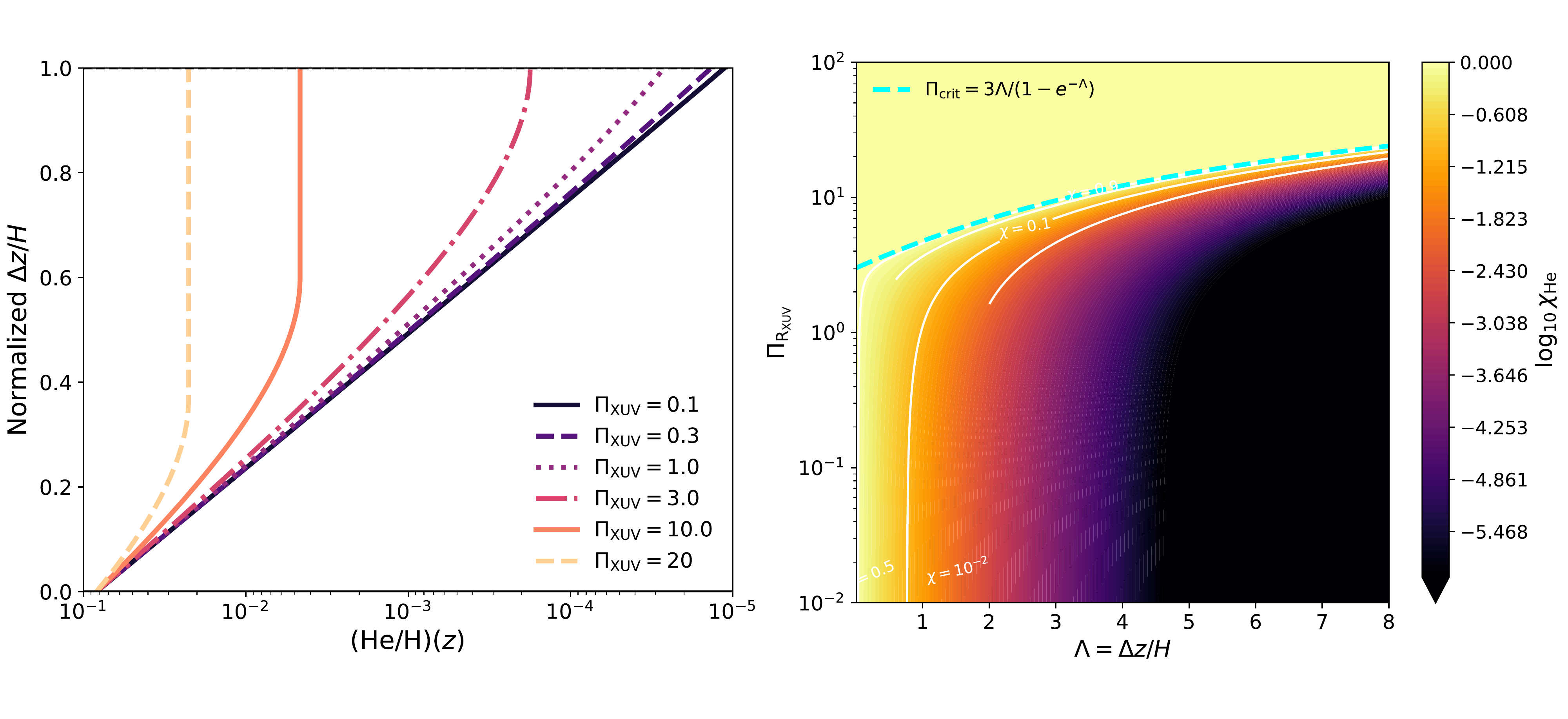}
\caption{Left: Ratio across the transition region for a fixed atmospheric extent $\Lambda=3$
and different values of the transport efficiency at the XUV heating layer, $\Pi_{\rm XUV}$. The vertical coordinate is normalised such that 0 corresponds to the homopause and 1 to the wind base layer. In the compositionally coupled limit, the profiles approach the primordial homopause abundance asymptotically. Right: Transport phase diagram for the helium retention factor $\chi_{\text{He}}$
within the homopause-wind base region. The horizontal axis shows the dimensionless atmospheric extent
while the vertical axis shows the transport efficiency at the wind base,
The colour scale represents the predicted helium retention factor. with bright regions corresponding to compositionally coupled atmospheres and dark regions indicating strong helium depletion due to molecular diffusion. White contours show constant values of $\chi_{\rm He}$. The cyan dashed curve marks the analytic transition condition $\Pi_{\rm crit}$ which separates diffusion-dominated and advection-dominated transport regimes.}
\label{fig:figure2}
\end{figure*}

Figure~\ref{fig:figure2} illustrates the dependence of the helium
retention factor on the two controlling dimensionless parameters of the
problem, namely the atmospheric extent $\Lambda$ and the transport
efficiency $\Pi_{\rm XUV}$. The figure provides a general map of the
transport--retention regimes expected in primordial H/He atmospheres,
independent of any particular planetary system.

Several general trends emerge immediately. In the limit
$\Pi_{\rm XUV}\ll1$, vertical transport is inefficient compared with
molecular diffusion and helium becomes strongly depleted between the
homopause and the XUV heating layer. In this regime, $\chi_{\rm He} \rightarrow e^{-3\Lambda}$,
such that the retention factor depends only on the geometric extent of
the transition region. As $\Lambda$ increases, helium must diffuse
across a larger number of scale heights before reaching the escaping
flow, resulting in progressively stronger depletion.

Conversely, in the limit $\Pi_{\rm XUV}\gg1$, upward transport
overwhelms diffusive separation and the atmosphere remains
compositionally coupled. The helium abundance at the base of the wind
approaches its deep atmospheric value, $\chi_{\rm He}
\rightarrow 1$, indicating nearly complete retention of the primordial He/H ratio. The transition between these two regimes occurs when transport and
diffusion become comparable. As shown in Figure~\ref{fig:figure2}, the
critical transport efficiency required to maintain a well-mixed
atmosphere increases rapidly with atmospheric extent. Compact
atmospheres ($\Lambda\lesssim0.1$) remain helium-rich even for modest
transport efficiencies, whereas highly extended atmospheres
($\Lambda\gtrsim1$) require substantially stronger upward transport to
prevent diffusive helium depletion.

The parameter $\Lambda$ may be interpreted as the number of scale
heights separating the homopause from the base of the escaping wind.
Values $\Lambda\sim0.05-0.1$ correspond to relatively compact
thermospheres in which the wind base layer lies only a short
distance above the homopause. In contrast, values
$\Lambda\sim1-5$ describe highly extended upper atmospheres
characteristic of strongly irradiated sub-Neptunes and hot gas giants,
where the escaping flow originates many scale heights above the
well-mixed lower atmosphere. Figure~\ref{fig:figure2} therefore
suggests that helium depletion should be most pronounced in planets
combining extended upper atmospheres with weak upward transport,
whereas strongly escaping atmospheres are expected to preserve nearly
primordial He/H ratios throughout the transition region.

\section{Application to observed exoplanets}

\subsection{Helium retention and observability}

The retention factor derived in the previous sections quantifies the
helium abundance available at the base of the escaping wind. However,
the observed He\,{\sc i}\,10830\,\AA\ absorption is not determined by
composition alone. The strength of the signal depends on a variety of
additional processes, including the atmospheric mass-loss rate,
thermospheric temperature structure, ionisation balance, metastable
helium production efficiency, and radiative transfer effects.
Consequently, the retention factor should not be interpreted as a
direct predictor of the observed equivalent width. Instead, it
represents a boundary condition that determines how much helium is
available to participate in the upper-atmospheric processes responsible
for producing the observable signal.

\begin{figure*}
\centering
\includegraphics[scale=0.52, trim=0cm 1cm 0cm 0cm, clip]{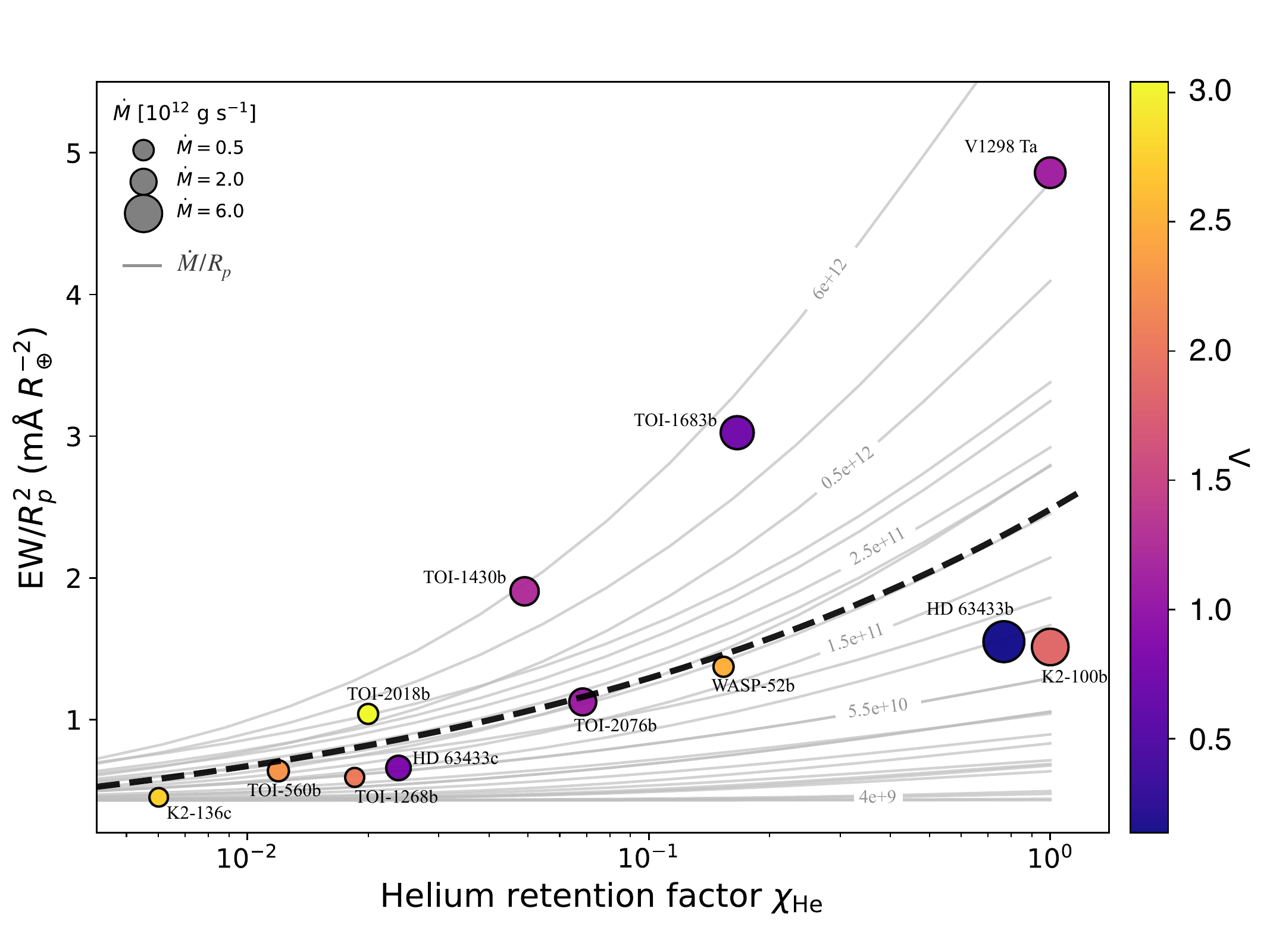}
\caption{Equivalent width of the He,{\sc i},10830,\AA\ triplet normalized by
$R_p^2$ as a function of the helium retention factor $\chi_{\rm He}$. Grey curves show a parametric grid of \texttt{p-winds} calculations spanning different prescribed helium abundances at the base of the escaping flow and different mass-loss rates. The \texttt{p-winds} calculations do not include transport-driven helium fractionation, but instead predict the He,{\sc i},10830,\AA\ absorption for a prescribed lower-boundary helium abundance. Coloured circles correspond to the sample
of helium-observed exoplanets from \citet{Allan2025}, with horizontal
positions given by the retention factors predicted by the analytic
model and vertical positions by the observed equivalent widths. Marker
colour indicates the atmospheric extent parameter $\Lambda$, while
marker size scales with the estimated mass-loss rate. }
\label{fig:corner}
\end{figure*}

To explore the observational consequences of helium retention, we performed a parametric grid of \texttt{p-winds}\footnote{\url{https://github.com/ladsantos/p-winds}}  (\citet{DosSantos2022}) calculations spanning a range of helium abundances prescribed at the base of the escaping atmosphere and a range of mass-loss rates. These calculations are not coupled to the analytic transport model. Instead, they quantify the expected He,{\sc i},10830,\AA\ equivalent width as a function of the lower-boundary helium abundance for different atmospheric escape rates.  The grey curves in Figure~\ref{fig:corner} show the predicted equivalent width
normalized by $R_p^2$ for a range of mass-loss rates and helium
abundances at the base of the escaping flow. To remove the first-order geometric dependence of the transit signal on planetary size, all equivalent widths are normalised by $R_p^2$. The motivation for this normalisation is discussed in Appendix~A. Each curve corresponds to a fixed hydrodynamic solution, while the horizontal coordinate represents the helium retention factor, which may be interpreted as a
proxy for the He/H abundance supplied to the wind. 

Several general trends emerge immediately. For all mass-loss rates
considered, the predicted helium absorption increases monotonically
with increasing retention factor. This behaviour is expected because
larger values of $\chi_{\rm He}$ imply a larger helium reservoir
available for populating the metastable triplet state. At fixed
retention factor, the equivalent width also increases with mass-loss
rate, reflecting the larger column densities present in stronger
planetary winds. The figure therefore demonstrates that helium
observability is controlled by at least two independent ingredients:
the amount of helium reaching the wind and the strength of the escaping
flow itself. The purpose of these calculations is not to reproduce the observations
of individual planets, but rather to provide a physical interpretation
of how transport-driven helium depletion modifies the expected helium
signal. In this framework, the retention factor acts as a supply term,
while the remaining atmospheric and radiative processes determine how
efficiently that helium reservoir is converted into observable
absorption.

\subsection{Predicted retention states of observed helium planets}

We now apply the analytic model to the sample of twelve helium-observed
planets analysed by \citet{Allan2025}. This sample is particularly
valuable because it combines high-quality He\,{\sc i}\,10830\,\AA\
observations with detailed hydrodynamic simulations extending from the
hydrostatic atmosphere into the escaping flow. These simulations
therefore provide the information required to estimate the two
controlling parameters of our analytic framework, namely the
atmospheric extent $\Lambda$ and the transport efficiency $\Pi$.

For each planet, in the sample, we estimated $\Lambda$ and $\Pi$ from the published
planetary and atmospheric properties (\citet{Allan2025}) and subsequently computed the
helium retention factor using Equation~(\ref{eq:chi_he_final}). The
planetary properties, adopted atmospheric boundary conditions, and
derived transport parameters used in these calculations are summarised
in Table~\ref{tab:planet_parameters}. The resulting helium retention
factors are shown as colored circles in Figure~\ref{fig:corner}. The
horizontal position of each point indicates the predicted helium
retention factor, while the vertical position corresponds to the
observed equivalent width normalised by $R_p^2$. The colour scale
denotes the atmospheric extent parameter $\Lambda$, while the marker
size scales with the estimated mass-loss rate. Unlike the grey \texttt{p-winds} curves, which represent theoretical calculations for prescribed helium abundances and mass-loss rates, the
coloured circles correspond to actual exoplanets. Their horizontal
positions are therefore predictions of the analytic transport model,
whereas their vertical positions are direct observational
measurements.

It is important to note that neither \texttt{p-winds} nor the hydrodynamic simulations of \citet{Allan2025} self-consistently include transport-driven diffusive separation below the base of the planetary wind. Instead, both prescribe the helium abundance entering the escaping flow as a lower-boundary condition. Consequently, the comparison presented in Figure~\ref{fig:corner} is not intended as a validation of the analytic transport-retention theory itself, but rather as a comparison of the predicted He,{\sc i},10830,\AA\ absorption once the lower-boundary helium abundance has been specified. In the present work, this lower-boundary abundance is provided by the analytic retention factor derived above. The assumptions underlying the transport-retention derivation are therefore not part of this comparison, since the forward models are used only to compute the observable for a prescribed lower-boundary helium abundance.

The planets span a remarkably broad range of predicted retention
states, from nearly complete retention
($\chi_{\rm He}\simeq1$) to severe depletion
($\chi_{\rm He}\simeq4\times10^{-3}$). This corresponds to a variation
of more than two orders of magnitude in the helium abundance available
at the base of the escaping wind. Such diversity indicates that
transport and diffusive separation alone can substantially modify the
composition supplied to the escaping atmosphere. Although the observed planets display a wide range of equivalent
widths, Figure~\ref{fig:corner} shows that many occupy distinct
transport-retention states. Planets exhibiting comparable helium
signals may therefore possess very different helium abundances at the
base of their winds. In this sense, helium observability is determined
not only by the strength of atmospheric escape, but also by the amount
of helium that survives transport through the transition region between
the homopause and the escaping flow.

The principal implication is that helium depletion does not necessarily
require intrinsically helium-poor atmospheres, chemical evolution, or
the presence of secondary atmospheric compositions. Instead,
substantial depletion can arise naturally from the competition between
upward transport and molecular separation.

\begin{table*}
\centering
\caption{Planetary properties and adopted model parameters used for the planet sample shown in Figure~\ref{fig:corner}. The planetary masses ($M_p$), radii ($R_p$), orbital separations ($a$), atmospheric base radii ($R_{\rm base}$), base temperatures ($T_{\rm base}$), and atmospheric escape rates ($\dot{M}$) are adopted from the hydrodynamic models of \citet{Allan2025}. The terminal velocity ($v_{\rm term}$) is the wind velocity at a distance of $10R_p$. The dimensionless transport parameter $\Pi_{\rm He,H}$ is computed from the adopted planetary and atmospheric properties, and the predicted helium retention factor $\chi_{\rm He}$ is obtained from Equation~(\ref{eq:chi_he_final}).}
\label{tab:planet_parameters}
\begin{tabular}{lcccccccccc}
\hline\hline
Planet &
$M_p$ ($M_\oplus$) &
$R_p$ ($R_\oplus$) &
$a$ (AU) &
$R_{\rm base}$ ($R_p$) &
$T_{\rm base}$ (K) &
$v_{\rm term}$ (m\,s$^{-1}$) &
$\Lambda$ &
$\dot{M}$ ($10^{12}$ g\,s$^{-1}$) &
$\Pi_{\rm He,H}$ &
$\chi_{\rm He}$ \\
\hline
Wasp-52b    & 137.9 & 14.0 & 0.027 & 1.25 & 987  & 67 & 1.601 & 2.00 & 3.668  & 0.154 \\
V1298 Tau c & 26.7  & 5.5  & 0.082 & 1.43 & 829  & 31 & 1.046 & 6.50 & 4.838  & 1.000 \\
TOI-1268b   & 96.3  & 9.1  & 0.072 & 1.16 & 676  & 23 & 1.417 & 0.50 & 0.346  & 0.019 \\
TOI-1683b   & 8.0   & 2.3  & 0.036 & 1.51 & 666  & 27 & 0.894 & 3.40 & 1.500  & 0.166 \\
K2-100b     & 21.8  & 3.9  & 0.030 & 1.65 & 1375 & 34 & 1.343 & 7.60 & 5.453  & 1.000 \\
TOI-1430b   & 7.0   & 2.1  & 0.070 & 1.49 & 604  & 21 & 1.101 & 1.90 & 0.432  & 0.049 \\
TOI-2018b   & 9.2   & 2.3  & 0.061 & 1.32 & 479  & 11 & 1.817 & 0.30 & 1.836  & 0.020 \\
TOI-2076b   & 9.0   & 2.6  & 0.063 & 1.49 & 650  & 23 & 1.035 & 2.00 & 0.656  & 0.068 \\
HD~63433b   & 5.3   & 2.2  & 0.072 & 1.77 & 729  & 26 & 0.654 & 5.70 & 3.533  & 0.767 \\
HD~63433c   & 7.3   & 2.7  & 0.146 & 1.49 & 512  & 20 & 0.930 & 1.40 & -1.567 & 0.024 \\
TOI-560b    & 15.9  & 2.8  & 0.060 & 1.26 & 541  & 19 & 1.511 & 0.60 & 0.137  & 0.012 \\
K2-136c     & 18.1  & 3.0  & 0.110 & 1.18 & 402  & 8  & 1.706 & 0.06 & 0.006  & 0.006 \\
\hline
\end{tabular}
\end{table*}
\section{Discussion}
\label{sec:discussion}

The principal result of this work is that the helium abundance supplied
to an escaping atmosphere is not necessarily equal to the helium
abundance of the deep atmosphere. Instead, the composition at the base
of the hydrodynamic flow is controlled by the competition between
upward transport and molecular separation within the transition region
connecting the homopause to the planetary wind base. The resulting
retention factor, $\chi_{\rm He}$, provides a simple analytic
description of this competition and predicts that the helium abundance
available to the escaping flow may vary by more than two orders of
magnitude even among planets that retain primordial H/He envelopes.

Our results complement previous multi-fluid hydrodynamic numerical simulations. Those studies demonstrated that helium can become depleted relative to hydrogen in escaping atmospheres through diffusive fractionation. The analytic solution derived here isolates the transport occurring between the homopause and the base of the wind, providing an explicit dependence of the helium retention factor on the atmospheric extent and transport efficiency. Despite their different methodologies, both approaches predict the same qualitative behaviour: weak upward transport relative to molecular diffusion promotes helium depletion, whereas efficient hydrodynamic transport maintains compositions closer to the primordial atmospheric abundance. In addition to reproducing this qualitative behaviour, the analytic solution identifies the governing dimensionless parameters and provides simple scaling relations that can be evaluated without performing full multi-fluid simulations.

Diffusive separation is expected to continue its effect within the escaping flow itself. The present theory does not assume that fractionation ceases at the base of the wind. Rather, it isolates the transport processes that determine the composition supplied to the escaping flow. In this sense, the helium retention factor derived here provides the lower boundary condition for the escaping atmosphere.  The observed composition of the escaping atmosphere should therefore be regarded as the cumulative result of transport through the homopause-wind transition region and any subsequent fractionation within the hydrodynamic outflow.

Low He/H ratios inferred from helium observations
are often discussed in terms of atmospheric evolution, preferential
escape, or transitions away from primordial composition (E.g., \citet{Kobayashi2026}). Our analysis shows that similar levels of depletion can arise naturally from transport physics operating between the deep atmosphere and the
escaping flow, even when the atmosphere as a whole retains a nebular
composition. In this sense, the composition of the escaping wind should
not be assumed to represent the composition of the bulk atmosphere.

The numerical calculations presented here further demonstrate that
helium retention acts primarily as a supply term for the observable
He\,{\sc i}\,10830\,\AA\ signal. Increasing $\chi_{\rm He}$ increases
the helium reservoir available to the upper atmosphere and therefore
enhances the maximum achievable absorption signal. The observed
equivalent width, however, remains sensitive to additional processes,
including the mass-loss rate, thermospheric temperature structure,
ionisation balance, and radiative transfer effects. The retention
factor should therefore be viewed as a physically motivated boundary
condition for helium observability rather than as a direct predictor of
the observed absorption strength.

Application of the model to the helium-observed sample of
\citet{Allan2025} suggests that currently observed systems span nearly
the full range of retention states predicted by the analytic
framework, from almost complete retention
($\chi_{\rm He}\simeq1$) to severe depletion
($\chi_{\rm He}\simeq4\times10^{-3}$). This diversity implies that
planets exhibiting similar helium absorption strengths may nevertheless
possess substantially different helium abundances at the base of their
escaping atmospheres. Interpreting helium observations solely in terms
of escape strength may therefore overlook an important component of the
underlying physics. Our framework provides a simple connection between the
composition of the deep atmosphere and that of the escaping flow.
Although the present study focuses on primordial H/He atmospheres, the
same transport-retention formalism can be generalised to heavier
species and more complex atmospheric compositions. Future work should
combine the analytic solution with detailed multi-species hydrodynamic
simulations and larger helium observational samples to determine how
transport-driven fractionation shapes atmospheric evolution across the
broader exoplanet population.

\begin{acknowledgements}

\textit{Acknowledgments:}  The authors acknowledge financial support from the Severo Ochoa grant CEX2021-001131-S funded by MCIN/AEI/10.13039/501100011033 and  Ministerio de Ciencia e Innovación through the project PID2022-137241NB-C43.

\end{acknowledgements}

\appendix

\section{Geometric normalisation of the helium equivalent width}
\label{app:normalization}

The observed He\,{\sc i}\,10830\,\AA\ equivalent width is computed from the excess absorption relative to the nearby continuum, which comes from the planetary transit radius, $\sim R_p$. In order to focus our sample comparison only on the aeronomical physical processes, we seek a first-order normalisation factor to eliminate geometric effects of each signal. The choice of an appropriate normalisation depends on the characteristic size of the absorbing region. In classical transmission spectroscopy, the atmosphere is assumed to remain hydrostatic and confined to a thin shell above the planetary radius (E.g., \citet{Gkouvelis2026a}).  The absorbing region is confined to a thin annulus, $b_{\rm sat}=R_p+H$, where $H$ is the atmospheric scale height and $H\ll R_p$. In this limit, $(R_p+H)^2-R_p^2\simeq 2R_pH$, where the geometric factor is in this case $R_p$. 
Helium absorption in escaping atmospheres, in contrast,  is produced by an extended hydrodynamic outflow whose radial extent may reach several planetary radii  $ b_{\rm sat}\sim2R_p,\,3R_p,\,5R_p $.  A useful insight is provided by the transmission theory of planetary winds (\citet{Gkouvelis2026b}). In the optically thin regime, the chord optical depth scales as
\begin{equation}
\tau(b,\lambda)
\propto
\frac{\sigma(\lambda)\dot M}
     {b\,v(b)},
\end{equation}
where $\sigma(\lambda)$ is the absorption cross section, $\dot M$ is the atmospheric mass-loss rate, $b$ is the impact parameter, and $v(b)$ is the local wind velocity. The characteristic radius of the absorbing region may be estimated by defining a saturation radius $b_{\rm sat}$ such that$\tau(b_{\rm sat},\lambda)\sim 1.$ This condition yields the excess absorbing area is$A_{\rm abs}
\sim
\pi
\left(
b_{\rm sat}^2-R_p^2
\right)$, which implies
\begin{equation}
{\rm EW}
\propto
\frac{
b_{\rm sat}^2-R_p^2
}
{R_*^2}.
\end{equation}
The dominant geometric scaling is therefore no longer proportional to $R_pH$, but instead to the projected planetary area, $R_p^2$.  This provides a physical motivation for normalising the helium equivalent width and isolating the remaining dependence on atmospheric escape physics. Empirically, this interpretation is supported by both the numerical \texttt{p-winds} models and the observational sample considered in this work, which exhibit a substantially reduced scatter when the equivalent width is normalised by $R_p^2$. In Figure \ref{fig:corner}, we show both the numerical p-winds simulations, which smoothly combine at very low helium content at the base of the wind, and as we increase the content, the curves spread as $\dot{M}/R_p$. 

To first order, variations between planets enter through parameters such as the mass-loss rate, temperature, gravity, and composition, which modify the dimensionless extent of the absorbing atmosphere, $b_{\rm sat}/R_p$. The absolute projected area still contributes a factor $R_p^2$, leading to $
{\rm EW}
\propto
R_p^2
\,F
\left(
\chi_{\rm He},
\dot M,
T,
g,
\ldots
\right)$.

\bibliography{sample702}{}

\begin{thebibliography}{}
\expandafter\ifx\csname natexlab\endcsname\relax\def\natexlab#1{#1}\fi
\providecommand{\url}[1]{\href{#1}{#1}}
\providecommand{\dodoi}[1]{doi:~\href{http://doi.org/#1}{\nolinkurl{#1}}}
\providecommand{\doeprint}[1]{\href{http://ascl.net/#1}{\nolinkurl{http://ascl.net/#1}}}
\providecommand{\doarXiv}[1]{\href{https://arxiv.org/abs/#1}{\nolinkurl{https://arxiv.org/abs/#1}}}

\bibitem[{A.~P. {Allan} {\&} A.~A. {Vidotto}(2025){Allan} \&
  {Vidotto}}]{Allan2025}
{Allan}, A.~P., \& {Vidotto}, A.~A. 2025, \bibinfo{title}{{Helium escape
  signatures are generally strongest during younger ages but this age
  dependence is lost in the diversity of observed exoplanets},} \mnras, 539,
  2144, \dodoi{10.1093/mnras/staf566}

\bibitem[{P.~M. Banks {\&} G. Kockarts(1973)Banks \& Kockarts}]{Banks1973}
Banks, P.~M., \& Kockarts, G. 1973, Aeronomy (New York, NY, USA: Academic
  Press)

\bibitem[{L.~A. {Dos Santos} {et~al.}(2022){Dos Santos}, {Vidotto},
  {Vissapragada}, {Alam}, {Allart}, {Bourrier}, {Kirk}, {Seidel}, \&
  {Ehrenreich}}]{DosSantos2022}
{Dos Santos}, L.~A., {Vidotto}, A.~A., {Vissapragada}, S., {et~al.} 2022,
  \bibinfo{title}{{p-winds: An open-source Python code to model planetary
  outflows and upper atmospheres},} \aap, 659, A62,
  \dodoi{10.1051/0004-6361/202142038}

\bibitem[{L. {Gkouvelis}(2026{\natexlab{a}}){Gkouvelis}}]{Gkouvelis2026a}
{Gkouvelis}, L. 2026{\natexlab{a}}, \bibinfo{title}{{A Closed-form Analytical
  Theory of Nonisobaric Transmission Spectroscopy for Exoplanet Atmospheres},}
  \apj, 997, 307, \dodoi{10.3847/1538-4357/ae3246}

\bibitem[{L. {Gkouvelis}(2026{\natexlab{b}}){Gkouvelis}}]{Gkouvelis2026b}
{Gkouvelis}, L. 2026{\natexlab{b}}, \bibinfo{title}{{A theory of transmission
  spectroscopy of hydrodynamic outflows from planetary atmospheres:
  Spectral-line saturation and limits on mass-loss constraints},} \aap, 710,
  A62, \dodoi{10.1051/0004-6361/202658963}

\bibitem[{L. {Gkouvelis} {et~al.}(2025){Gkouvelis}, {Pozuelos}, {Drant},
  {Farhat}, {Tian}, \& {Ak{\i}n}}]{Gkouvelis2025}
{Gkouvelis}, L., {Pozuelos}, F.~J., {Drant}, T., {et~al.} 2025,
  \bibinfo{title}{{Interior redox state effects on the stability of secondary
  atmospheres and observational manifestations: LP 791-18 d as a case study for
  outgassing rocky exoplanets},} \aap, 699, A378,
  \dodoi{10.1051/0004-6361/202554192}

\bibitem[{D.~M. {Hunten} {et~al.}(1987){Hunten}, {Pepin}, \&
  {Walker}}]{Hunten1987}
{Hunten}, D.~M., {Pepin}, R.~O., \& {Walker}, J.~C.~G. 1987,
  \bibinfo{title}{{Mass fractionation in hydrodynamic escape},} \icarus, 69,
  532, \dodoi{10.1016/0019-1035(87)90022-4}

\bibitem[{J. {Kirk} {et~al.}(2020){Kirk}, {Alam}, {L{\'o}pez-Morales}, \&
  {Zeng}}]{Kirk2020}
{Kirk}, J., {Alam}, M.~K., {L{\'o}pez-Morales}, M., \& {Zeng}, L. 2020,
  \bibinfo{title}{{Confirmation of WASP-107b's Extended Helium Atmosphere with
  Keck II/NIRSPEC},} \aj, 159, 115, \dodoi{10.3847/1538-3881/ab6e66}

\bibitem[{I. {Kobayashi} {et~al.}(2026){Kobayashi}, {Kurokawa}, {Schaefer}, \&
  {Okuzumi}}]{Kobayashi2026}
{Kobayashi}, I., {Kurokawa}, H., {Schaefer}, L., \& {Okuzumi}, S. 2026,
  \bibinfo{title}{{Helium Depletion in Escaping Atmospheres of Sub-Neptunes: A
  Signature of Primary-to-secondary Transition},} \apj, 997, 110,
  \dodoi{10.3847/1538-4357/ae226b}

\bibitem[{M. {Lamp{\'o}n} {et~al.}(2023){Lamp{\'o}n}, {L{\'o}pez-Puertas},
  {Sanz-Forcada}, {Czesla}, {Nortmann}, {Casasayas-Barris}, {Orell-Miquel},
  {S{\'a}nchez-L{\'o}pez}, {Danielski}, {Pall{\'e}}, {Molaverdikhani},
  {Henning}, {Caballero}, {Amado}, {Quirrenbach}, {Reiners}, \&
  {Ribas}}]{Lampon2023}
{Lamp{\'o}n}, M., {L{\'o}pez-Puertas}, M., {Sanz-Forcada}, J., {et~al.} 2023,
  \bibinfo{title}{{Characterisation of the upper atmospheres of HAT-P-32 b,
  WASP-69 b, GJ 1214 b, and WASP-76 b through their He I triplet absorption},}
  \aap, 673, A140, \dodoi{10.1051/0004-6361/202245649}

\bibitem[{D.~C. {Linssen} {et~al.}(2022){Linssen}, {Oklop{\v{c}}i{\'c}}, \&
  {MacLeod}}]{Linssen2022}
{Linssen}, D.~C., {Oklop{\v{c}}i{\'c}}, A., \& {MacLeod}, M. 2022,
  \bibinfo{title}{{Constraining planetary mass-loss rates by simulating Parker
  wind profiles with Cloudy},} \aap, 667, A54,
  \dodoi{10.1051/0004-6361/202243830}

\bibitem[{R.~A. {Murray-Clay} {et~al.}(2009){Murray-Clay}, {Chiang}, \&
  {Murray}}]{Murray2009}
{Murray-Clay}, R.~A., {Chiang}, E.~I., \& {Murray}, N. 2009,
  \bibinfo{title}{{Atmospheric Escape From Hot Jupiters},} \apj, 693, 23,
  \dodoi{10.1088/0004-637X/693/1/23}

\bibitem[{L. {Nortmann} {et~al.}(2018){Nortmann}, {Pall{\'e}}, {Salz},
  {Sanz-Forcada}, {Nagel}, {Alonso-Floriano}, {Czesla}, {Yan}, {Chen},
  {Snellen}, {Zechmeister}, {Schmitt}, {L{\'o}pez-Puertas}, {Casasayas-Barris},
  {Bauer}, {Amado}, {Caballero}, {Dreizler}, {Henning}, {Lamp{\'o}n}, {Montes},
  {Molaverdikhani}, {Quirrenbach}, {Reiners}, {Ribas}, {S{\'a}nchez-L{\'o}pez},
  {Schneider}, \& {Zapatero Osorio}}]{Nortmann2018}
{Nortmann}, L., {Pall{\'e}}, E., {Salz}, M., {et~al.} 2018,
  \bibinfo{title}{{Ground-based detection of an extended helium atmosphere in
  the Saturn-mass exoplanet WASP-69b},} Science, 362, 1388,
  \dodoi{10.1126/science.aat5348}

\bibitem[{A. {Oklop{\v{c}}i{\'c}} {\&} C.~M. {Hirata}(2018){Oklop{\v{c}}i{\'c}}
  \& {Hirata}}]{Oklopvcic2018}
{Oklop{\v{c}}i{\'c}}, A., \& {Hirata}, C.~M. 2018, \bibinfo{title}{{A New
  Window into Escaping Exoplanet Atmospheres: 10830 {\r{A}} Line of Helium},}
  \apjl, 855, L11, \dodoi{10.3847/2041-8213/aaada9}

\bibitem[{J. {Orell-Miquel} {et~al.}(2024){Orell-Miquel}, {Murgas},
  {Pall{\'e}}, {Mallorqu{\'\i}n}, {L{\'o}pez-Puertas}, {Lamp{\'o}n},
  {Sanz-Forcada}, {Nortmann}, {Czesla}, {Nagel}, {Ribas}, {Stangret},
  {Livingston}, {Knudstrup}, {Albrecht}, {Carleo}, {Caballero}, {Dai},
  {Esparza-Borges}, {Fukui}, {Heng}, {Henning}, {Kagetani}, {Lesjak}, {de
  Leon}, {Montes}, {Morello}, {Narita}, {Quirrenbach}, {Amado}, {Reiners},
  {Schweitzer}, \& {Vico Linares}}]{Orell2024}
{Orell-Miquel}, J., {Murgas}, F., {Pall{\'e}}, E., {et~al.} 2024,
  \bibinfo{title}{{The MOPYS project: A survey of 70 planets in search of
  extended He I and H atmospheres: No evidence of enhanced evaporation in young
  planets},} \aap, 689, A179, \dodoi{10.1051/0004-6361/202449411}

\bibitem[{J.~E. {Owen}(2019){Owen}}]{Owen2019}
{Owen}, J.~E. 2019, \bibinfo{title}{{Atmospheric Escape and the Evolution of
  Close-In Exoplanets},} Annual Review of Earth and Planetary Sciences, 47, 67,
  \dodoi{10.1146/annurev-earth-053018-060246}

\bibitem[{J. {Sanz-Forcada} {et~al.}(2025){Sanz-Forcada}, {L{\'o}pez-Puertas},
  {Lamp{\'o}n}, {Czesla}, {Nortmann}, {Caballero}, {Zapatero Osorio}, {Amado},
  {Murgas}, {Orell-Miquel}, {Pall{\'e}}, {Quirrenbach}, {Reiners}, {Ribas},
  {S{\'a}nchez-L{\'o}pez}, \& {Solano}}]{Forcada2025}
{Sanz-Forcada}, J., {L{\'o}pez-Puertas}, M., {Lamp{\'o}n}, M., {et~al.} 2025,
  \bibinfo{title}{{Connection between planetary He I {\ensuremath{\lambda}}10
  830 {\r{A}} absorption and extreme-ultraviolet emission of planet-host
  stars},} \aap, 693, A285, \dodoi{10.1051/0004-6361/202451680}

\bibitem[{M. {Schulik} {\&} J.~E. {Owen}(2025){Schulik} \&
  {Owen}}]{Schulik2025}
{Schulik}, M., \& {Owen}, J.~E. 2025, \bibinfo{title}{{Using the helium triplet
  as a tracer of the physics of giant planet outflows},} \mnras, 542, 927,
  \dodoi{10.1093/mnras/staf775}

\bibitem[{J.~J. {Spake} {et~al.}(2018){Spake}, {Sing}, {Evans},
  {Oklop{\v{c}}i{\'c}}, {Bourrier}, {Kreidberg}, {Rackham}, {Irwin},
  {Ehrenreich}, {Wyttenbach}, {Wakeford}, {Zhou}, {Chubb}, {Nikolov}, {Goyal},
  {Henry}, {Williamson}, {Blumenthal}, {Anderson}, {Hellier}, {Charbonneau},
  {Udry}, \& {Madhusudhan}}]{Spake2018}
{Spake}, J.~J., {Sing}, D.~K., {Evans}, T.~M., {et~al.} 2018,
  \bibinfo{title}{{Helium in the eroding atmosphere of an exoplanet},} \nat,
  557, 68, \dodoi{10.1038/s41586-018-0067-5}

\bibitem[{A.~R. {Taylor} {et~al.}(2026){Taylor}, {Koskinen}, {Huang}, {Arfaux},
  \& {Lavvas}}]{Taylor2026}
{Taylor}, A.~R., {Koskinen}, T.~T., {Huang}, C., {Arfaux}, A., \& {Lavvas}, P.
  2026, \bibinfo{title}{{Helium Escape in Context: Comparative Signatures of
  Four Close-in Exoplanets},} \apj, 999, 214, \dodoi{10.3847/1538-4357/ae41b5}

\bibitem[{A.~J. {Watson} {et~al.}(1981){Watson}, {Donahue}, \&
  {Walker}}]{Watson1981}
{Watson}, A.~J., {Donahue}, T.~M., \& {Walker}, J.~C.~G. 1981,
  \bibinfo{title}{{The dynamics of a rapidly escaping atmosphere: Applications
  to the evolution of Earth and Venus},} \icarus, 48, 150,
  \dodoi{10.1016/0019-1035(81)90101-9}

\bibitem[{L. {Xing} {et~al.}(2023){Xing}, {Yan}, \& {Guo}}]{Xing2023}
{Xing}, L., {Yan}, D., \& {Guo}, J. 2023, \bibinfo{title}{{The Mass
  Fractionation of Helium in the Escaping Atmosphere of HD 209458b},} \apj,
  953, 166, \dodoi{10.3847/1538-4357/ace43f}

\bibitem[{R.~V. {Yelle}(2004{\natexlab{a}}){Yelle}}]{Yelle2004}
{Yelle}, R.~V. 2004{\natexlab{a}}, \bibinfo{title}{{Aeronomy of extra-solar
  giant planets at small orbital distances},} \icarus, 170, 167,
  \dodoi{10.1016/j.icarus.2004.02.008}

\bibitem[{M. {Zhang} {et~al.}(2022){Zhang}, {Knutson}, {Wang}, {Dai}, \&
  {Barrag{\'a}n}}]{Zhang2022}
{Zhang}, M., {Knutson}, H.~A., {Wang}, L., {Dai}, F., \& {Barrag{\'a}n}, O.
  2022, \bibinfo{title}{{Escaping Helium from TOI 560.01, a Young
  Mini-Neptune},} \aj, 163, 67, \dodoi{10.3847/1538-3881/ac3fa7}

\end{thebibliography}
\bibliographystyle{aasjournalv7.1}

\end{document}